\documentclass[]{interact}

\usepackage[natbibapa,nodoi]{apacite}
\usepackage[
    colorlinks,
    citecolor=blue,
    urlcolor=blue,
    linkcolor=blue
    ]{hyperref}

\usepackage{graphicx} % Include figure files
\usepackage{dcolumn}  % Align table columns on decimal point
\usepackage{bm} % bold math
\usepackage[export]{adjustbox} % multipanels

\begin{document}
\title{The Impact of Parenthood on labor Market Outcomes of Women and Men in Poland} 

\author{
\name{
Radost Waszkiewicz\textsuperscript{a}\thanks{CONTACT R. Waszkiewicz; email: radost.waszkiewicz@gmail.com} and Honorata Bogusz\textsuperscript{b}}
\affil{
\textsuperscript{a}Institute of Theoretical Physics, \\
Faculty of Physics, University of Warsaw, Pasteura 5, 02-093 Warsaw, Poland\\
\textsuperscript{b}Interdisciplinary Centre for labor Market and Family Dynamics, \\
Faculty of Economic Sciences, University of Warsaw, Długa 44/50, 00-241 Warsaw, Poland
}
}

\date{\today}

\maketitle

\begin{abstract}
We examine the gender gap in income in Poland in relation to parenthood status, employing the placebo event history method adapted to low-resolution data (Polish Generations and Gender Survey). Our analysis reveals anticipatory behavior in both women and men who expect to become parents. We observe a decrease of approximately 20 percent in mothers' income post-birth. In contrast, the income of fathers surpasses that of non-fathers both pre- and post-birth, suggesting that the fatherhood child premium may be primarily driven by selection. We note an increase (decrease) in hours worked for fathers (mothers). Finally, we compare the gender gaps in income and wages between women and men in the sample with those in a counterfactual scenario where the entire population is childless. Our findings indicate no statistically significant gender gaps in the counterfactual scenario, leading us to conclude that parenthood drives the gender gaps in income and wages in Poland.
\end{abstract}


%JEL Code: J13 - Fertility; Family Planning; Child Care; Children; Youth
%JEL Code: J16 - Economics of Gender; Non-labor Discrimination
%JEL Code: J31 - Wage Level and Structure; Wage Differentials by Skill, Training, Occupation, etc.

\section{Introduction}\label{introduction}

\begin{figure}[htbp]
    \centering
    \begin{tabular}{cc}
    \footnotesize A)
    \includegraphics[width=0.45\linewidth,valign=t]{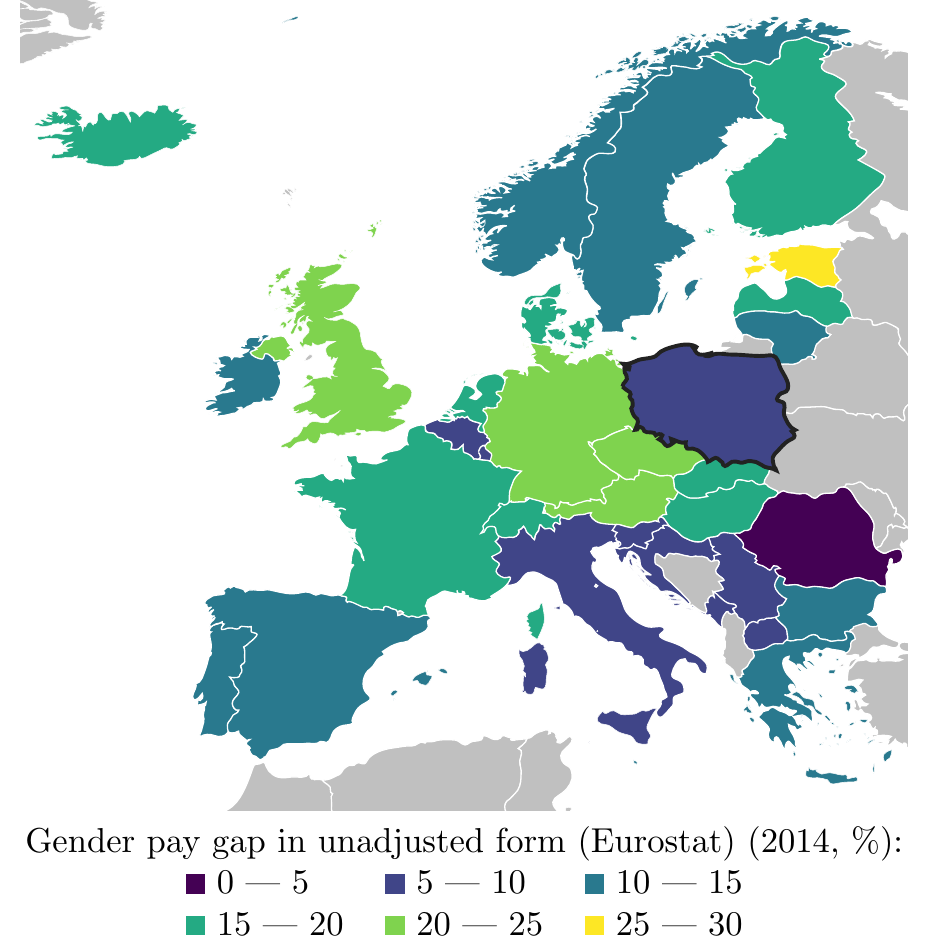} &
    \footnotesize B)
    \includegraphics[width=0.45\linewidth,valign=t]{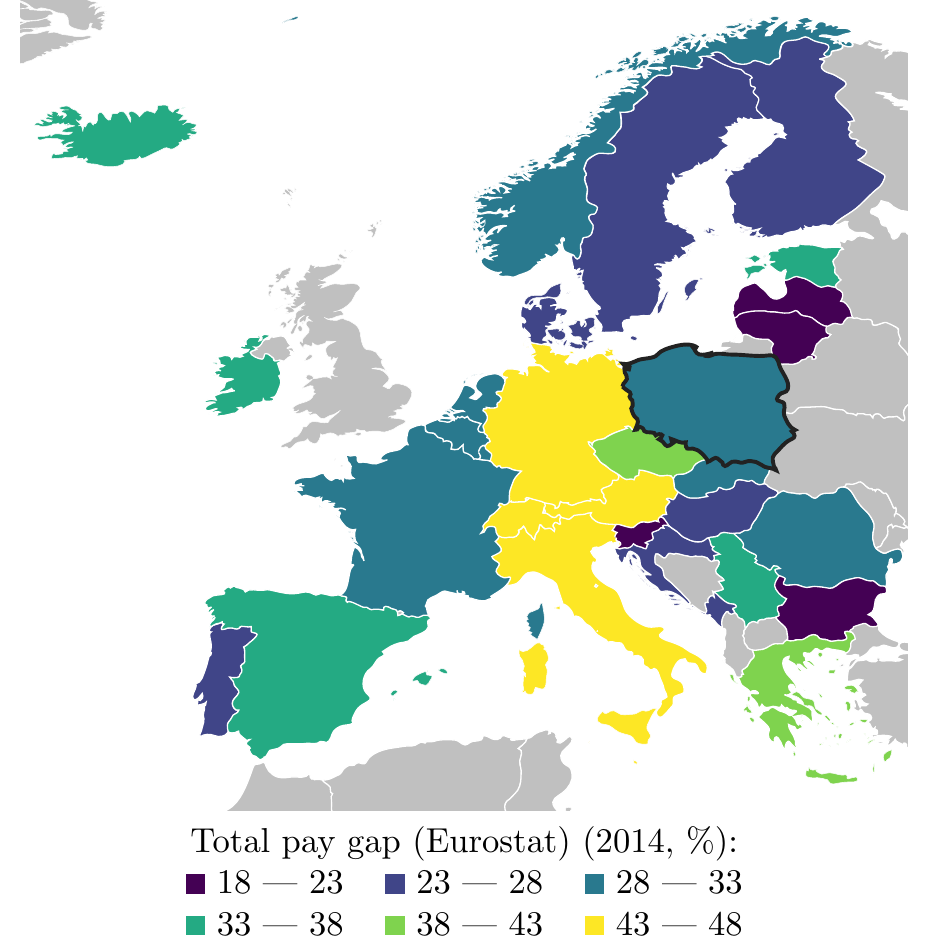}    
    \end{tabular}
    \caption{\textbf{A)} Unadjusted gender pay gap data from Eurostat show differences in hourly earnings. In this methodology countries like Romania, Italy and Poland share a very low (below 5 percent) value of GPG.
    \textbf{B)} Gender overall earnings gap from \cite{Eurostat2023}: \textit{The gender overall earnings gap is a synthetic indicator. It measures the impact of the three combined factors, namely: (1) the average hourly earnings, (2) the monthly average of the number of hours paid (before any adjustment for part-time work) and (3) the employment rate, on the average earnings of all women of working age - whether employed or not employed - compared to men.} Here, countries with low female employment rates, which record the lowest gender wage gaps, like Poland, Romania, or Italy, actually score higher gender overall earnings gaps than the more gender-egalitarian Scandinavian countries, for example.}
    \label{fig:maps_GPG}
\end{figure}

Poland boasts one of the lowest gender pay gaps in Europe (see Figure \ref{fig:maps_GPG}, Panel A). Additionally, it holds one of the highest proportions of women employed in STEM fields \citep{Eurostat2022} and in managerial positions \citep{Eurostat2021}. Nevertheless, Poland remains a socially conservative country, with approximately 92 percent of the population aged 16 and above identifying as Catholic \citep[][data for 2018]{GUS2022}. In recent years, women's rights have regressed, exemplified by the enactment of the abortion ban in January 2021, rendering Poland the sole EU country where abortion is illegal. This glaring disparity between the economic and social facets of gender equality in Poland serves as the cornerstone of this study. We argue that evaluating the gender gap in hourly wages, the conventional way for assessing differential economic outcomes between women and men \citep[e.g.,][]{Blau2017, Goldin2014}, proves inadequate within the Polish labor market. 
The principal reason lies in Poland's notably low female labor force participation rates (Figure \ref{fig:labor_force_participation}). 
Consequently, framing the gender gap in economic outcomes solely through hourly wages, as frequent in economic studies, overlooks nearly half of Poland's working-age female population. 
Moreover, juxtaposing such calculations with those of Western European counterparts may inaccurately portray the economic equality seemingly experienced by women in Poland.
Instead, we advocate for the use of total personal monthly income after taxes. 
Applying a modified placebo event method proposed by \cite{Kleven2019}, we investigate the income trajectories of women and men in relation to their parental status. 
This approach quantifies the role of parenthood in shaping differential labor market outcomes for women and men, as evidenced in numerous economic studies \citep[e.g.,][]{Goldin2021, Kleven2019}.

While childless women often experience increasing earnings over their lifetimes, new mothers typically adjust their labor market behavior and encounter declines in earnings, income, employment, and hours worked—a phenomenon commonly referred to as the \emph{child penalty} \citep{Blau2017}. 
These penalties are intertwined with patriarchal gender norms \citep{Matysiak2021, Cukrowska2020, Dominguez2022}, although it remains unclear whether such norms mediate or condition the declines in economic outcomes experienced by mothers.
Initially, explanations for the causes of child penalties centered on disparities in human capital/labor market advantages among workers \citep[e.g.,][]{Altonji1999}, but these theories were debunked, alongside explanations based on the biological costs of childbearing and access to policies that facilitate the combination of paid work and caregiving responsibilities \citep{Andresen2022, Kleven2022, Kleven2021_2}. 
This suggests that the origins of child penalties may lie elsewhere. 
While no definitive research offers a clear explanation for the causes of child penalties, an ongoing study by \cite{Kleven2023} has demonstrated that they tend to manifest with increases in a country's development and wealth.

In contrast to mothers, fathers experience a \emph{child premium}, typically defined by an increase in earnings following the birth of a child, which is not observed among childless men. 
This premium arises from an uptick in hours worked, along with a propensity to take on additional jobs or pursue career changes leading to better-paying positions or promotions \citep{Baranowska2022, Cukrowska2020}. 
Previous studies primarily attributed these phenomena to the Beckerian specialization hypothesis, rooted in neoclassical theories of time allocation \citep{Becker1965}. 
However, with the growing acceptance of women's employment and the narrowing of the gender gap in education, men's labor market advantages have substantially diminished, if not vanished entirely \citep{Baranowska2022}. 
Recent research suggests that the relatively favorable labor market conditions of fathers compared to childless men may be influenced by normative perceptions of fatherhood \citep{Hodges2010, Baranowska2022}. 
In societies where masculinity is closely associated with the breadwinner role, fathers are perceived as ideal workers willing to commit to long hours with minimal disruptions from family obligations. 
Furthermore, the \emph{fatherhood child premium} may be attributed to a selection mechanism, whereby men who are successful in the labor market and possess greater resources are more inclined to have children \citep{Baranowska2022}.

This paper offers two main contributions. First, by concentrating on Poland, we augment the gender-economic literature concerning understudied Eastern European nations. 
Moreover, Poland presents an intriguing case study in its own right due to the apparent gender neutrality of its labor market, as mentioned earlier, and the largely ambiguous gender norms for women (see Section \ref{sec_background}). 
We build upon two previous notable contributions concerning the gender gap in economic outcomes in Poland, both of which examined the gender pay gap \citep{Cukrowska2016, Goraus2017}. 
Through a decomposition approach, \cite{Cukrowska2016} revealed that the gender pay gap in Poland is primarily influenced by the fatherhood child premium, while the motherhood penalty plays a smaller yet significant role. 
\cite{Goraus2017} demonstrated that, after adjusting for employment selection, the gender pay gap doubles in magnitude compared to its unadjusted form. 
Although the study by \cite{Goraus2017} employed a different methodology than ours, its substantive contribution bears the closest resemblance to ours in assessing the gender gap in income, i.e., incorporating non-working women when investigating gendered labor market outcomes.

Similar to the aforementioned papers, much of the existing literature on this topic has relied methodologically on a decomposition approach (refer to \cite{Goraus2017} for a discussion), typically employing the Oaxaca-Blinder decomposition, occasionally supplemented by adjustments for selection into employment. 
Following the pioneering work of \cite{Kleven2019}, the literature has transitioned towards the event study approach \citep[e.g.,][]{Andresen2022, Kleven2022, Kleven2021_2}. 
The specific \textit{placebo event method} shares similarities with the difference-in-difference design. 
However, this approach has seen limited application in studies conducted in countries lacking high-quality administrative birth registers accessible to researchers\footnote{With the exception of the ongoing research by \cite{Kleven2022}, where the placebo event method is adapted for use with cross-sectional data. However, this adaptation differs from the one presented in our contribution.}. 
This limitation may stem from the fact that, in the original approach of \cite{Kleven2019}, counterfactual births are assigned to non-parents through numerical simulation, introducing computer-generated noise into the estimates, which poses challenges with small sample sizes. 
Our second contribution to the literature is methodological: we modify the placebo event method by deriving an analytical solution for the numerical experiment (see Section \ref{sec_methods}), significantly reducing noise and enabling placebo event studies on low-dimensional data. 
We illustrate the modified method using the Generations and Gender Survey (2010 and 2014\footnote{Given that we only have data for 2010 and 2014 at our disposal, we refer to data for the year 2014 when discussing additional descriptive statistics throughout the paper.}), the sole available Polish micro-level dataset containing information on both births (including parity) and income.

\section{Background}\label{sec_background}

\begin{figure}[htb]
    \centering
    \includegraphics[width=0.8\linewidth]{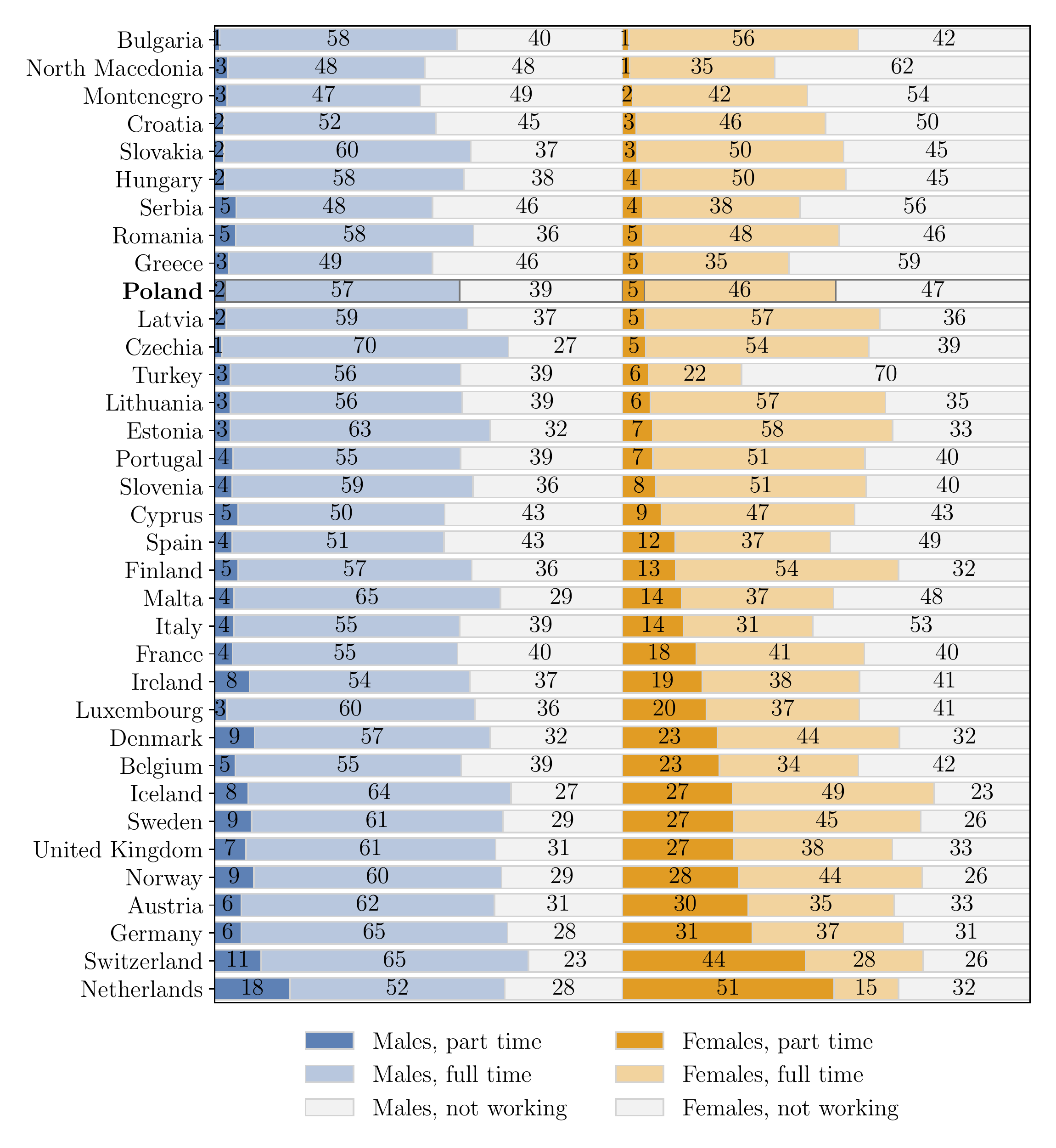}
    \caption{A cross-country comparison of labor force status in percent, 2014 (labor market data: EU-LFS aggregates; Population by age group: Eurostat). Poland shows very small shares of part-time work in both men and women compared to countries like Germany or United Kingdom.}
    \label{fig:labor_force_participation}
\end{figure}

Our focus in this contribution centers on Poland, the largest post-Soviet economy situated in Central Europe. Poland presents an intriguing case for studying child penalties due to its ambiguous gender norms. On one hand, both women (51 percent in 2014, 65 percent in 2022) and men (59 percent in 2014, 77 percent in 2022) exhibit relatively low employment rates compared to Western European countries (Figure \ref{fig:labor_force_participation}, \cite{EurostatEmplRates}).
However, the prevalence of part-time work is among the lowest in Europe (Figure \ref{fig:labor_force_participation}): in 2014, only 5 percent of women of working age engaged in part-time employment, in contrast to 31 percent in Germany, for instance. 
For men, the corresponding figures were 2 percent and 6 percent. 
The disparities between women in Poland and Western Europe are particularly striking and suggest that women in Poland predominantly work either full-time or not at all. 
Normatively, both women and men are expected to hold full-time jobs \citep{Matysiak2008, Reilly2014}, a tradition that may trace back to communist times when every adult was obligated to work full-time. 
However, the communist regime did not entirely depart from the "male production, female reproduction" model and instead adopted a "dual earner-female double burden" model \citep{Fidelis2010}. 
This perception of gender roles still persists in Poland today. 
If a small child is present at home, women are often expected to exit the workforce until the child attends kindergarten or primary school; otherwise, they are urged to swiftly secure employment and contribute to the household income \citep{Matysiak2011}. 
However, the proportion of women's contributions to household chores diminishes as their share of income toward the household budget increases, except in couples with gender attitudes that prioritize men's work \citep{Magda2023}.

\begin{figure}[htbp]
    \centering
    \begin{tabular}{cc}
    \footnotesize A)
    \includegraphics[width=0.45\linewidth,valign=t]{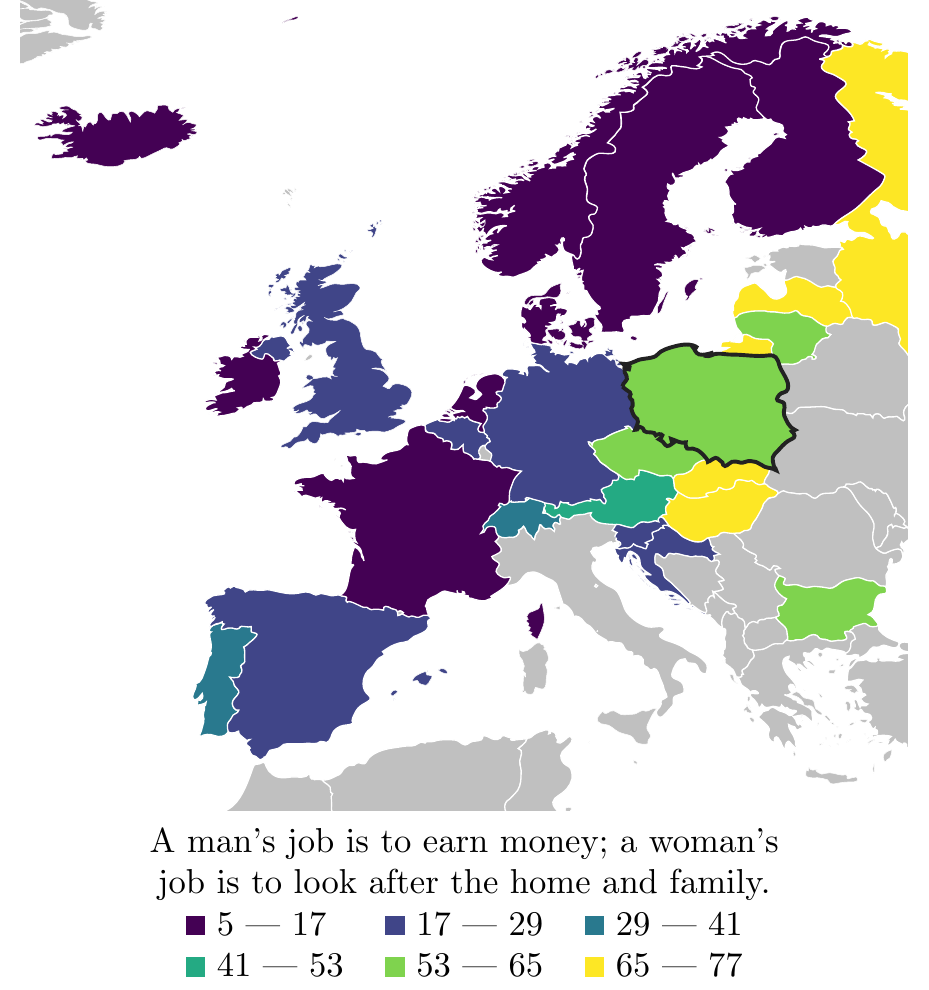} &
    \footnotesize B)
    \includegraphics[width=0.45\linewidth,valign=t]{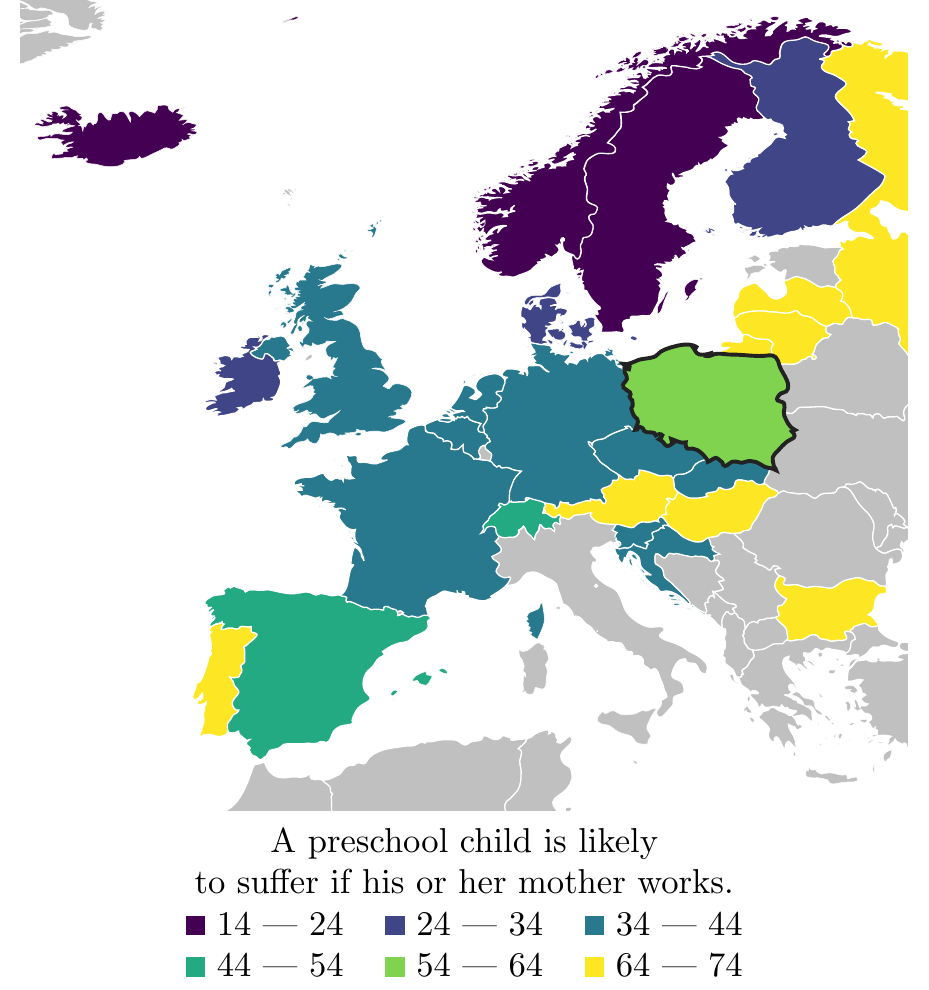}    
    \end{tabular}
    \caption{Percent of affirmative answers to two questions selected from International Social Survey Programme: Family and Changing Gender Roles \citep{ISSP}. Data for 2012.}
    \label{fig:maps_attitudes}
\end{figure}

The state reinforces the "dual earner-female double burden" model with "implicit familialism," which places families under significant strain \citep{Javornik2014}. 
The availability of institutionalized childcare remains limited \citep{Szelewa2008, Saxonberg2007}, and fathers are entitled to only two weeks of paternity leave (with a take-up rate of 55 percent, \cite{Ojcostwo2021}). 
While paid maternity and parental leaves extend up to 12 months in total, they can theoretically be shared by both parents, but only if the mother meets the eligibility criteria (i.e., employed under a work contract). However, merely 1 percent of fathers avail themselves of this option \citep{Ojcostwo2021}. 
These deeply ingrained gendered patterns are evident in Polish attitudes toward gender roles (see Figure \ref{fig:maps_attitudes}).

These phenomena are likely at least partially influenced by cultural norms deeply rooted in Catholicism. 
Nearly the entire adult population of Poland identifies as Catholic \citep[][data for 2018]{GUS2022}. 
The Polish Catholic Church actively promotes a narrative of "gender ideology" that upholds traditional (heteronormative) gender roles and emphasizes female subservience \citep{Szelewa2021, Szwed2017}. 
Following World War II and during the early communist era, the church opposed female employment \citep{Fidelis2010}. 
This perspective shifted somewhat with the acknowledgment that individuals at the lower rungs of society face economic imperatives, necessitating both partners to work \citep{Fidelis2010}. 
Presently, while the church does not fundamentally object to female employment, it underscores that a woman's primary role is that of a homemaker and caregiver \citep{Szwed2017}. 
Moreover, it opposes legislation aimed at enhancing reproductive rights or safeguarding victims of domestic abuse, among other measures, presenting itself as a defender of the traditional (heteronormative) Polish family, perceived to be under threat from Western "moral corruption" \citep{Szwed2017}. 
The church has been widely recognized as a primary impediment to the progress of the feminist movement in Poland, compared to other countries with similar levels of economic development \citep[e.g.,][]{Graff2007, Bystydzienski2001}, and is regarded as the primary political force responsible for Poland having the strictest abortion law in Europe \citep{Szelewa2016}.

The dynamics of the Polish labor market arguably contribute to the gendered nature of economic outcomes. 
Following the economic transition, Poland experienced an unemployment rate of 20 percent. 
However, despite challenges such as the Great Recession, the COVID-19 pandemic, and increased automation, unemployment has steadily declined, reaching a record low of 5 percent in 2022 (\cite{GUS_bezrobocie}, compared to 12 percent in 2014).
Correspondingly, the employment rate has increased, including the employment of women and older workers, although women still face underemployment \citep{EurostatEmplRates}. 
Nonetheless, low unemployment does not necessarily correlate with high job quality. 
Between 2012 and 2016, Poland had the highest proportion of temporary employment in the EU \citep{Lewandowski2023}. 
Polish workers, on average, work longer hours than their counterparts in most OECD countries \citep{OECD_hours_worked}. 
Moreover, Poland exhibits one of the highest rates of self-employment among EU countries—19.7 percent compared to the EU average of 14.5 percent in 2021 (21.4 percent and 16.4 percent, respectively, in 2014, \cite{OECD_self_empl}). 
Meanwhile, labor market protections in Poland are relatively low. 
The total compensation of employees (including wages, unemployment benefits, job creation programs, etc.) as a percentage of GDP ranks among the lowest in developed countries—47 percent compared to 62 percent in Germany and 58 percent in the United States \citep{UNECE_labour}. 
Real wages in Poland increased by 87 percent between 2000 and 2021, and earnings inequality decreased during this period, though it remains higher than in most European countries \citep{Lewandowski2023}. 
Trade union density is similar to that of Germany but is declining \citep{OECD_unions}. 
However, trade union density levels are comparable for women and men \citep{OECD_unions}.

In summary, the labor market and family formation dynamics in Poland likely perpetuate the gendered division of paid and unpaid labor within different-sex couples. 
The state's limited support for parents means that raising a child is predominantly viewed as a private responsibility. 
While full-time employment for women is normalized, they are also expected to prioritize child-rearing and shoulder the burden of household chores. 
The labor market is characterized by a high degree of precariousness, and wages tend to be relatively low, factors that may undermine family formation \citep{Alderotti2021}. 
These conditions are likely to contribute to gendered economic outcomes once a couple becomes parents.

\section{Data}\label{sec_methods_data}

We utilize both waves of the Polish Generations and Gender Survey (GGS), conducted in 2010 and 2014, respectively. 
The wave-matched database comprises 12,294 respondents, with 61 percent being women. 
The variables of interest include the respondent's age, the birth year of the first child, self-reported total post-tax income, and hours worked.

We categorized respondents as belonging to the childless group if they reported having no children by the second wave of the survey. 
The childlessness rate is approximately 15 percent for both women and men. 
Respondents with children were allocated to the mothers/fathers group if the year of their child's birth was known. 
Whenever possible, we used the specific month and day of the child's birth; otherwise, the date was set to the first day/month of the respective year.

The total personal income of each respondent was computed using the following questions from the first wave of the survey, in order of priority: the exact value of income, income bands, and no income flag. 
If income was specified using income bands (consisting of 13 possible bands), the midpoint of the band was assigned as the respondent's income, except for the "10,000+" band where 10,000 PLN was used as the midpoint (this applied to 15 cases in the dataset). Among respondents, 72 percent provided the exact value of their income, 16 percent provided income in bands, 3 percent indicated having no income, 7 percent refused to answer, and the remaining data was missing.

Data on total hours worked were computed by summing the responses to questions about the number of hours in the primary job and the number of hours worked in additional jobs. For individuals reporting no income in the GGS, both total income and hours worked were set to zero.

% Income data
% exact                 0.722567
% estimated             0.160015
% unavailable (98.0)    0.071766
% no_income             0.027688
% unavailable (97.0)    0.010279
% unavailable (-2.0)    0.006482
% parental_leave        0.001204

\section{Methodology}\label{sec_methods}

\begin{figure}[htb]
    \centering
    \includegraphics[width=.9\linewidth]{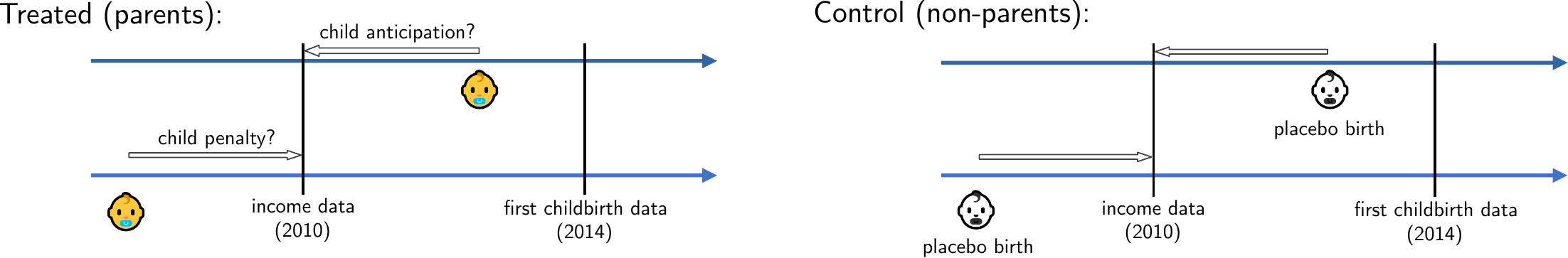}
    \caption{Pictorial representation of the method used.
    Data from the first wave of GGS (2010) are used to measure income while first childbirth year is picked from the second wave (2014). 
    We use birth events in years 2010-2014 to assess anticipatory behavior of parents and earlier births to measure child penalties. These are compared with a control group of non-parents who are age-matched to parents using a modified \emph{placebo event} method.}
    \label{fig:method_diagram}
\end{figure}

Most measures of labor market outcomes, such as income, are highly time-dependent, influenced by both age and the time elapsed since childbirth for parents.
This presents a challenge when assessing their dependence on parenthood. 
Consequently, a simple quantification of these effects is typically achieved by either calculating the difference between parents and non-parents \citep[e.g.,][]{Cukrowska2016, Blau2017, Bergsvik2019} or determining the difference between pre-child and post-child outcomes \citep[e.g.,][]{Angelov2016, Andresen2022, Kleven2019, Cortes2020}, but not both simultaneously.

Further insight can be gained from a difference-in-difference setting, provided that the pre-child behaviors of the reference and treatment groups are comparable. 
This condition holds true, for instance, in Denmark, as demonstrated by \cite{Kleven2019}. 
However, as we illustrate in Section \ref{sec_results}, this is not the case for Poland. 
To apply a difference-in-difference approach, the reference population must be matched with the treatment population by age.

The method employed by \cite{Kleven2019} involved generating a "placebo birth event" for childless individuals by randomly drawing the age at first birth from an appropriate distribution. 
The resulting dataset could be analyzed in a similar manner to the dataset of parents, enabling the quantification of a typical trajectory of, for instance, income given the number of years prior to or following the birth event (whether placebo or genuine).

The method, as utilized by \cite{Kleven2019}, which relies on numerical simulation to assign placebo births to childless individuals, tends to increase the noise in the placebo sample unnecessarily. 
Although measurement uncertainty is typically negligible in population-wide datasets, it can become prohibitively large with smaller samples. 
Since we have no particular preference for one sample of age at first placebo birth over another drawn from the same distribution, we could, in theory, repeat the numerical experiment numerous times. 
Consequently, we could obtain estimates that are entirely independent of the idiosyncrasies of the computer-drawn sample.
By computing (analytically) the expected value of the placebo group outcome measure under repeated re-randomization of the placebo births, we completely eliminate the noise arising from the numerical randomization procedure. 
Consequently, we decrease the standard error by a factor of approximately $\sqrt{M}$, where $M$ represents the number of bins in the age at first birth distribution (in our case, yearly bins, resulting in a reduction by a factor of approximately 5; for monthly bins, the reduction is by a factor of approximately 12). 
Similar to \cite{Kleven2019}, we assume that the age at first birth follows a lognormal distribution, which also fits well with our empirical data.

To obtain the \emph{placebo-event}-centered trajectory $Z_\tau$ of an outcome of interest (where $\tau$ represents the time since the event), one simply computes an appropriately weighted average of the outcome of interest $Y_t$ in age groups $t$ with the weights determined by the probability mass function $p(t)$ of the age at the event
\begin{equation}
    Z_\tau = \left. \left(\sum_t p(t-\tau) Y_t \right) \middle/ \left( \sum_t p(t-\tau) \right) \right.,
    \label{eqn:convolve}
\end{equation}
where the normalization factor in the denominator arises from the age range available in the dataset. 
Since the transformation \eqref{eqn:convolve} is a linear map on the $Y_t$ vector of age group means (each with an independent measurement error), the computation of the variance of each of the $Z_\tau$ variables is straightforward and done with a similar expression (unfortunately the covariance $Z_\tau$ is not diagonal). 
In contrast to the method employed by \cite{Kleven2019}, we report the absolute values of income instead of the differences relative to the value just before the birth event. 
This enables us to graphically compare pre-event trajectories between the treatment and the control group, including a constant offset which would be eliminated by the normalizing procedure.

Given that both the time since first birth and age increase over time, distinguishing between cohort and event effects can pose a challenge. 
When we examine values proximate to the event, the time elapsed since the event becomes a more significant variable.
Conversely, for larger values of lag, the cohort of the individual can be predicted with considerable accuracy based on the time elapsed since birth (all births under consideration happen within 4 years). 
The timescale of our analysis is therefore determined by the standard deviation of the age at first birth, which in our dataset amounts to $4.58$ years. 
The concept of proximity to or distance from the event should be measured against this standard deviation. Consequently, data for lags smaller than 10 years can be interpreted akin to a pseudo-panel, while greater caution is needed for larger lags.

After investigating the time-dependent trajectories, we compute the value of the gender wage gap and the gender gap in income in the observed scenario and in the counterfactual one, where no one has any children. 
The counterfactual values of the variable of interest $Y$ (income or wage) for an individual $i$ of gender $g$ and age $\textnormal{age}_i$ result from a regression of the form
\begin{equation}
        Y^{g,c}_{i} = \sum_k \beta_k^{g,c} \mathbb{I}[ k =  \textnormal{age}_i] + \varepsilon_i^{g,c},
    \label{eqn:non_parents_estimation}
\end{equation}
where $c$ is a child dummy variable. 
Due to the small size of our dataset, the age variable was rounded to the nearest multiple of 5. 
The regression parameters from equation \eqref{eqn:non_parents_estimation} are used to compute the counterfactual income $C^{g}_{i}$ for each individual in the counterfactual scenario where the entire population is childless as
\begin{equation}
    C^{g}_{i} = \hat{Y}^{g,c=0}_i.
\end{equation}
A simple average over $i$ is then computed to get a single value of income (or wage) in the counterfactual scenario for both genders
\begin{equation}
    \bar{C}^g = \frac{1}{n} \sum_i C_i^g.
    \label{eqn:average}
\end{equation}
We then compare the counterfactual outcomes of women and men with the observed ones, thereby assessing the impact of parenthood on gender gaps in income and wage.

\section{Results}\label{sec_results}

\begin{figure}[htb]
    \centering
    \begin{tabular}{cc}
         \footnotesize A) 
         \includegraphics[width=0.42\linewidth,valign=t]{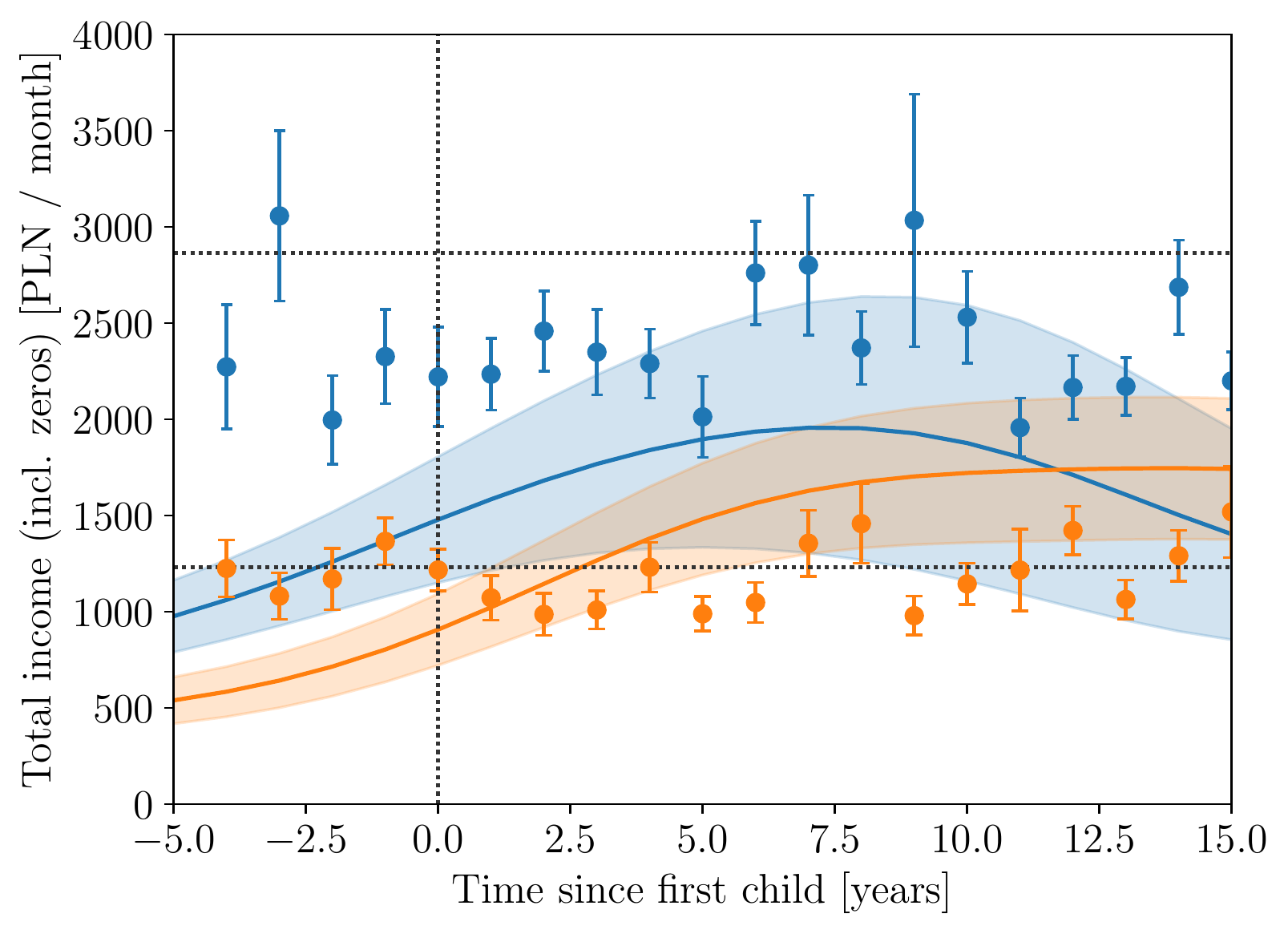} &  
         \footnotesize B) 
         \includegraphics[width=0.42\linewidth,valign=t]{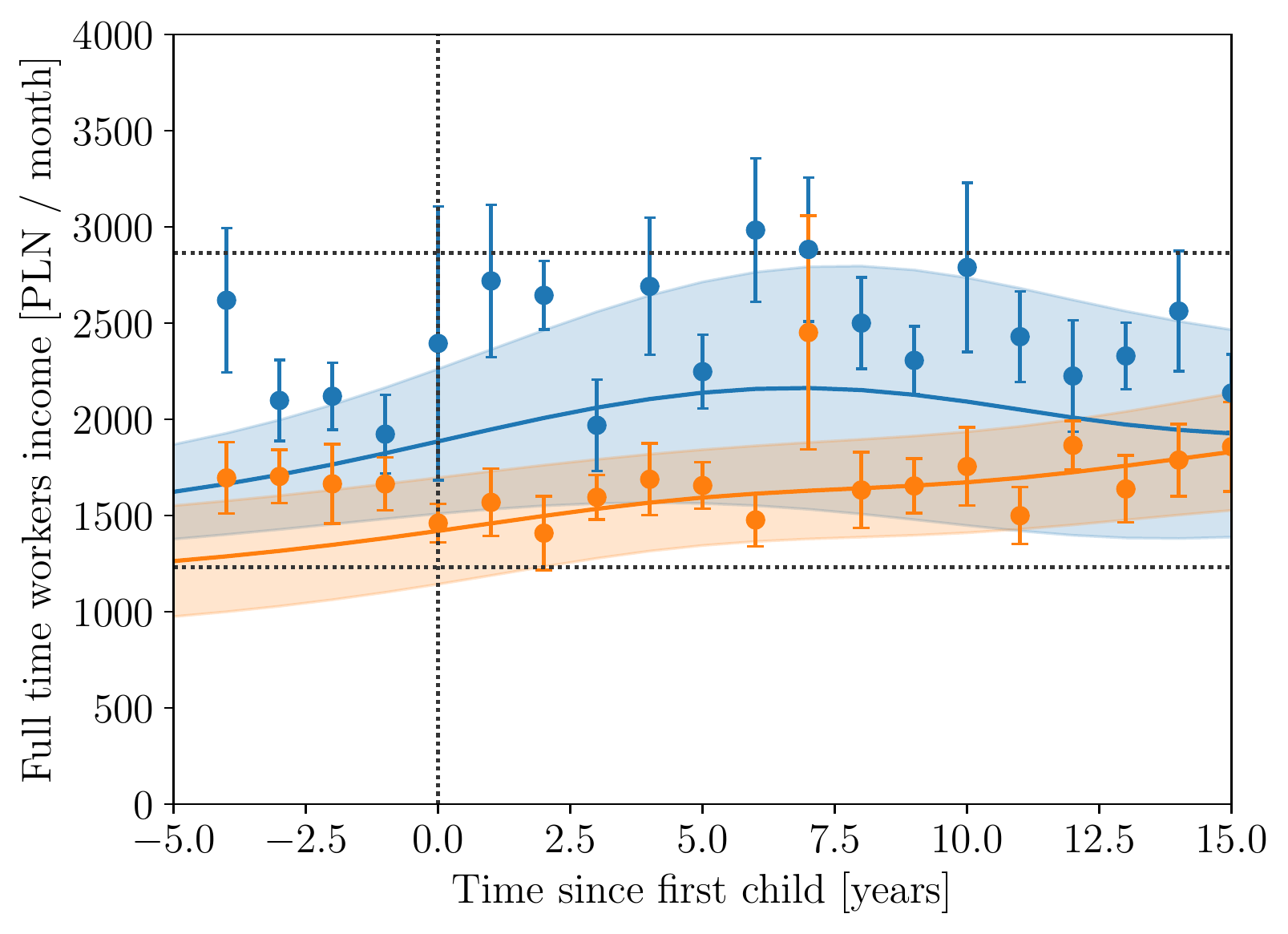}\\
         \multicolumn{2}{c}{
         \includegraphics[
         width=.7\linewidth,
         valign=t,
         clip=true,
         trim=0 4cm 0 4cm]{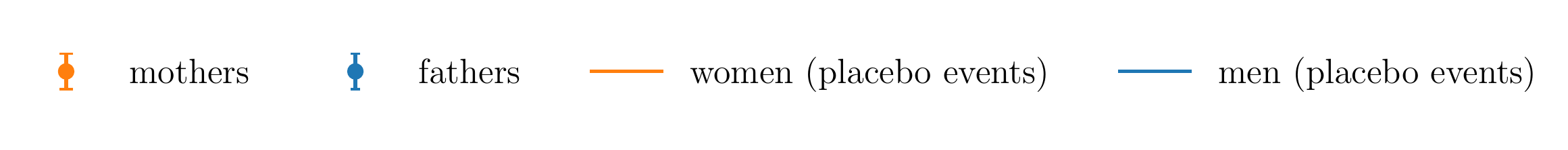}
         }
    \end{tabular}
    \caption{\textbf{A)}
    Mean of individual income of mothers and fathers (post tax, including individual transfers and individuals with no income and excluding household transfers) as a function of time since first childbirth, compared with the control groups of childless individuals estimated with the placebo event method. The horizontal lines show the minimum wage post tax (2014) and the mean wage of full-time (40 hr / week) workers in companies employing more than 9 workers (2014, tax adjusted).
    \textbf{B)}
    Analogous estimations to Panel A calculated for individuals working full-time.}
    \label{fig:total_income}
\end{figure}

\begin{figure}[htbp]
    \centering
    \begin{tabular}{cc}
         \footnotesize A) 
         \includegraphics[width=0.42\linewidth,valign=t]{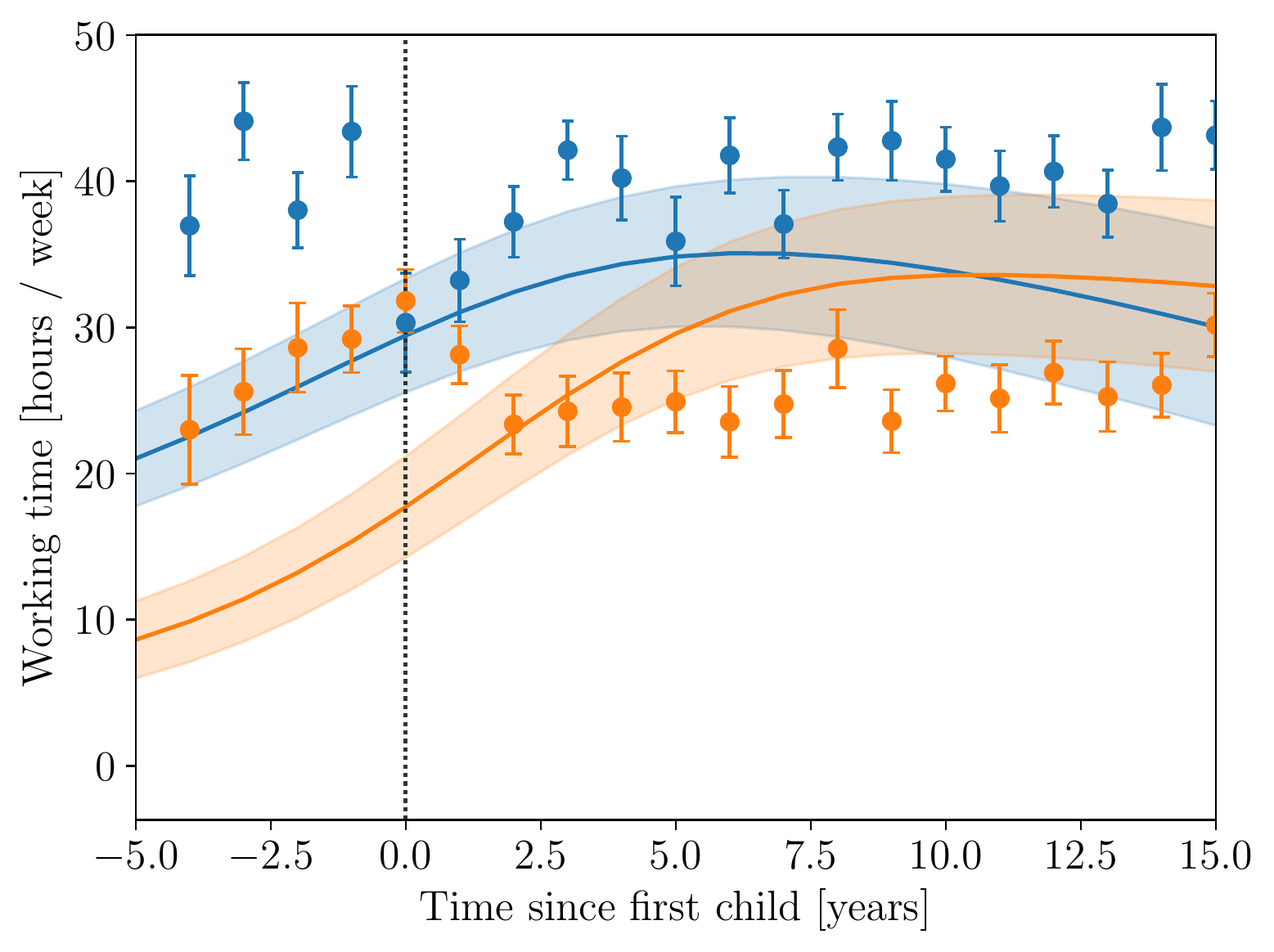} &  
         \footnotesize B) 
         \includegraphics[width=0.42\linewidth,valign=t]{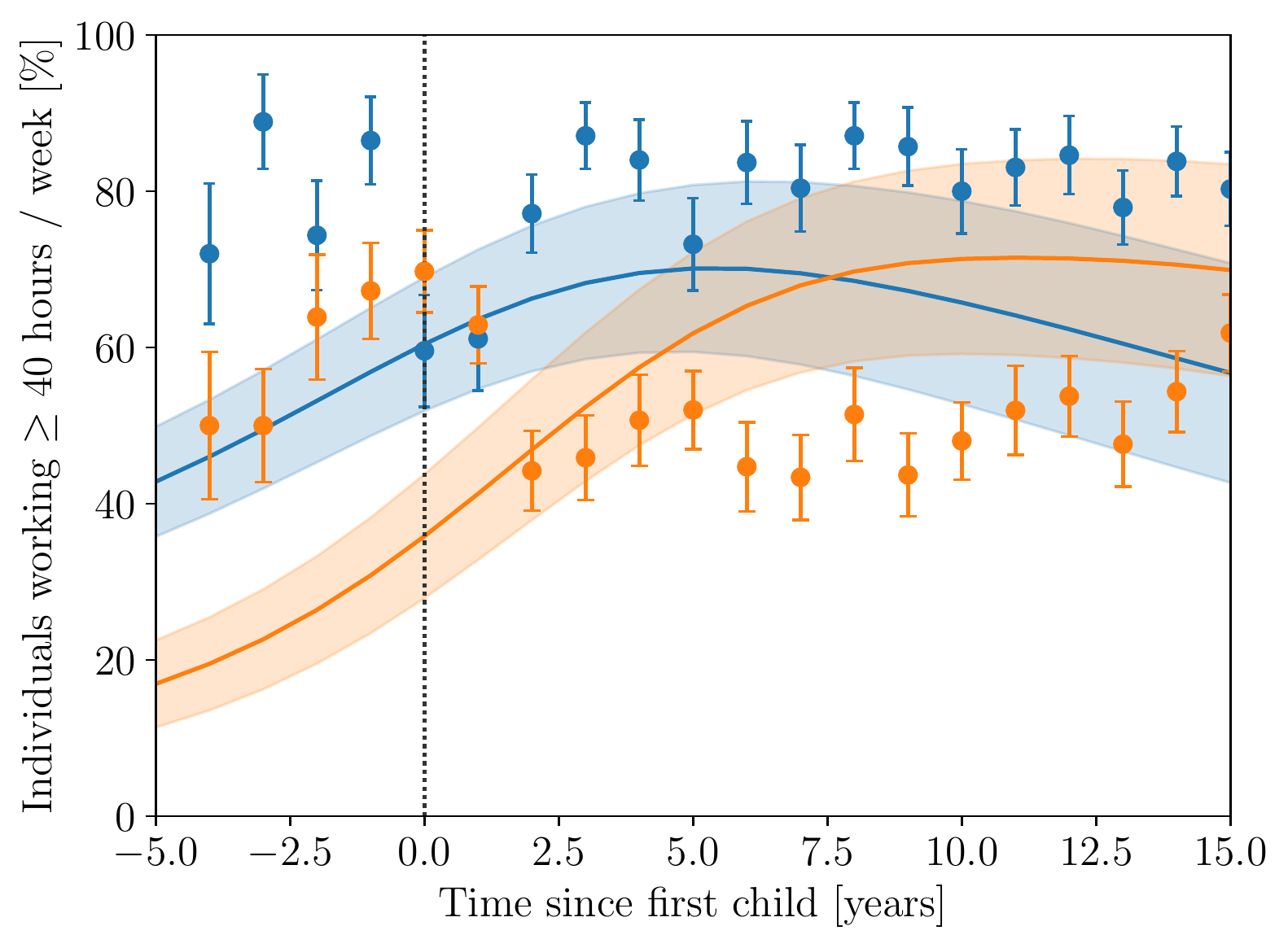}\\
         \multicolumn{2}{c}{
         \includegraphics[
         width=.7\linewidth,
         valign=t,
         clip=true,
         trim=0 4cm 0 4cm]{figures/placebo_legend.pdf}
         }
    \end{tabular}
    \caption{\textbf{A)}
    Mean number of hours worked as a function of time since first child compared with the control group of childless individuals estimated with the placebo event method (including the non-working population). The unexpectedly high number of hours worked for mothers in the first year past birth is an artifact of the questionnaire's design: women on maternity leave are asked about number of hours in the last job before taking the leave.
    \textbf{B)}
    Percent of population working at least 40 hours a week.}
    \label{fig:hours_worked}
\end{figure}

\begin{figure}[htbp]
    \centering 
    \setlength\tabcolsep{0pt}
    \begin{tabular}{cc}
    \footnotesize A) 
    \includegraphics[width=0.45\linewidth,valign=t]{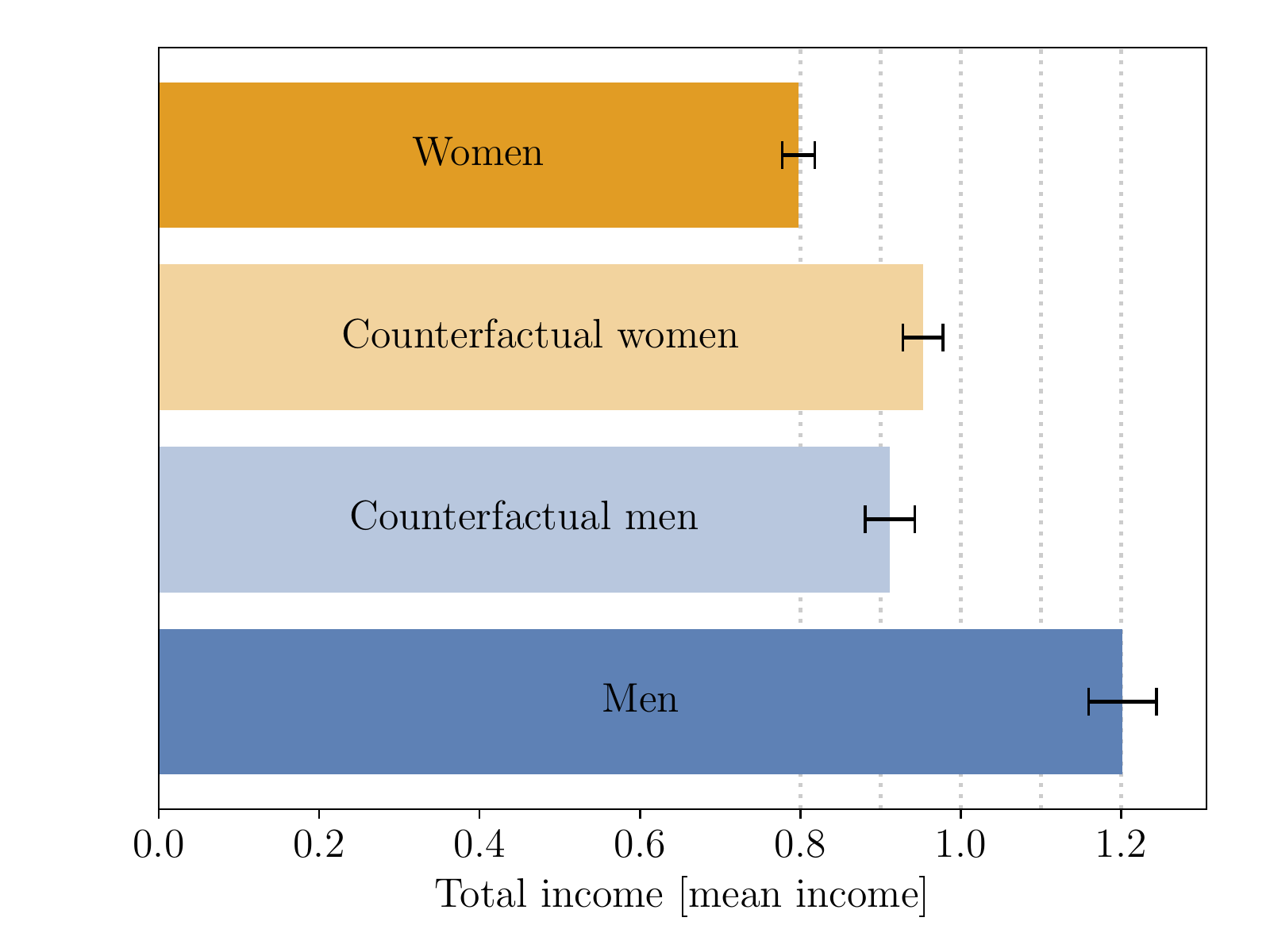} &
    \footnotesize B) 
    \includegraphics[width=0.45\linewidth,valign=t]{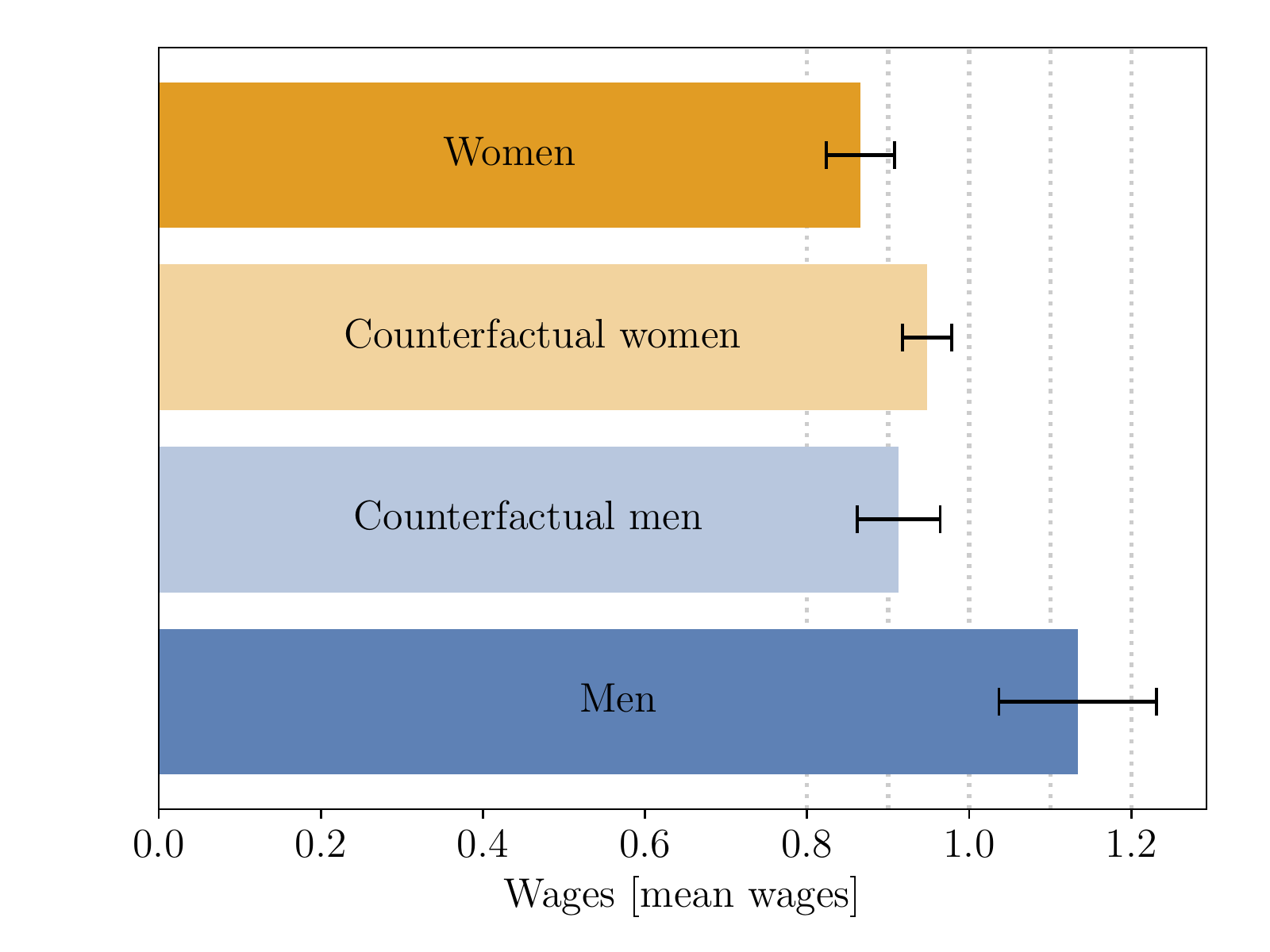}
    \end{tabular}
    \caption{\textbf{A)} Total income and total counterfactual income as a multiple of the real mean total income with confidence bands determined by bootstrap ($1\sigma$ CI of 50 rounds). 
    Incomes taken from 2010 survey and child data from 2014 survey (hence the adjustment includes anticipatory behaviors).
    \textbf{B)} Hourly wages of people working 40 hours a week as a multiple of the mean value in that group ($1\sigma$ CI of 50 bootstrap rounds). 
    Incomes taken from 2010 survey and child data from 2014 survey (hence the adjustment includes anticipatory behaviors).}
    \label{fig:counterfactuals}
\end{figure}

Similarly to the analysis of \cite{Kleven2019}, mothers fall below the control group after birth. However, unlike in Denmark, they return to the trend in about 15 years (Figure~\ref{fig:total_income}, Panel A). In contrast to Denmark, it is not true that future mothers and fathers follow the same path as the counterfactual parents -- they typically earn more in total, and this effect persists permanently after birth for fathers. 

A further issue with the assumption that both mothers and fathers, as well as the respective control groups, follow the same trajectories before birth is that fathers are typically older than mothers at the moment of the first childbirth, and they are at a different stage of the typical income trajectory. This alone would result in a total income differential between the control groups at the moment of birth of nearly 100 percent based on the placebo-event estimates (Figure~\ref{fig:total_income}, Panel A).

The deviations in total income between control groups and treated groups almost disappear when we restrict the dataset to only workers working 40 hours a week (the standard and most popular contract in Poland comprises 40 hours a week), as shown in Figure~\ref{fig:total_income}, Panel B. The anticipatory behavior is also largely diminished, and both the counterfactual trends and parental observations eventually converge to a similar value.

These observations are consistent with the research of \cite{Cukrowska2016}, who investigated the gender wage gap of full-time workers. 
They found that women incur a child penalty of about 2 percent, while men gain about a 7 percent premium, effects of a magnitude comparable to the error bands of our estimates. 
Hourly wage rates (specifically of full-time workers) have often been a primary variable of interest in studies of the motherhood penalty \citep[e.g.,][]{Cukrowska2020, Blau2017, Goldin2014}. 
However, in the case of Poland, it seems that parenthood's impact on earnings through that channel is small, albeit persistent over a long time.

The anticipatory behavior is most visible when comparing mean values of the total number of hours worked per week (see Figure~\ref{fig:hours_worked}, Panel A). 
Here, future mothers work nearly 50 percent more hours than the control group just before giving birth, only to fall below the trend in the later years (but eventually rejoin the trend after about 15 years). 
At the same time, fathers work more hours (about 30 percent) than their control group, and this effect appears to be permanent. 
A very similar pattern emerges when analyzing the share of individuals working at least 40 hours a week (Figure \ref{fig:hours_worked}, Panel B).

Finally, we compare the income and wages (Figure \ref{fig:counterfactuals}, Panel A and B, respectively) of the women and men in the GGS with the counterfactual scenario where the entire population is childless. The real gender gap in income is around 33 percent, while the counterfactual gap is not statistically significant. 
Similarly, the gender wage gap of full-time workers is around 20 percent, while the counterfactual gap is not statistically significant.

\section{Discussion}\label{sec_discussion}

We conclude that parenthood in Poland mainly has a significant impact on labor force participation and the willingness to work more hours than the standard contract. 
First, concerning women, we find that the earnings penalty is transient and not permanent, unlike in Denmark, and that this effect is almost entirely mediated by the mean number of hours worked. 
The fact that mothers return to their pre-birth income trajectory after 15 years might be driven by the normalization of full-time employment for women in Poland. 
Additionally, families enjoy lower support from the state than in Denmark, and they might face greater economic necessity. 
Consequently, even if women in Poland experience birth-related work disruptions, which can be especially long (until the child goes to primary school), they usually return to full-time employment.

Second, concerning men, we find that fathers consistently work more than the childless control group, both before and after childbirth. 
This shows that it is entirely possible to have a fatherhood premium without a significant jump at the birth event, likely due to a selection mechanism in which more affluent men decide to have children.

Our analysis of the income and hours worked trajectories uncovers a large degree of anticipatory behavior unseen in similar studies conducted in Western European setups. 
This anticipatory behavior might stem from the precariousness of the Polish labor market and low state support for families, which, combined, place the burden of childbearing and child-rearing on individuals. 
In such circumstances, future parents are likely to prepare economically for the arrival of an offspring. 
Additionally, this anticipatory behavior puts into question the formal difference-in-difference design, since some of the anticipatory effects can be masked by normalizing to $\tau = -1$ (as in the study of \cite{Kleven2019}, for example).

Our comparison of the gender gaps in income and wages between women and men in the GGS sample and the counterfactual scenario uncovers that both gaps are driven by parenthood. 
However, our analysis of the income trajectories shows that these effects stem from differences in labor market participation and hours worked, rather than wage rates. 
Thus, our results suggest that more focus is needed on the number of working hours and labor force participation in general when assessing the impact of parenthood on the economic outcomes of women and men. 
A general decrease in the number of hours worked by women in Poland is similar to the cases in Western Europe but is not universal worldwide or even in Europe, with opposite effects observed, for example, in Belarus \citep{Ganguli2014}.

\renewcommand{\BRetrievedFrom}{}%
\bibliographystyle{apacite}
\bibliography{sources}

\section*{Appendix A: Descriptive Statistcs}

\end{document}